\documentstyle[12pt,epsf]{article}
\catcode`\@=11
\long\def\@makefntext#1{
\protect\noindent \hbox to 3.2pt {\hskip-.9pt  
$^{{\ninerm\@thefnmark}}$\hfil}#1\hfill}		

\def\@makefnmark{\hbox to 0pt{$^{\@thefnmark}$\hss}}  
	
\def\ps@myheadings{\let\@mkboth\@gobbletwo
\def\@oddhead{\hbox{}
\rightmark\hfil\ninerm\thepage}   
\def\@oddfoot{}\def\@evenhead{\ninerm\thepage\hfil
\leftmark\hbox{}}\def\@evenfoot{}
\def\sectionmark##1{}\def\subsectionmark##1{}}

\renewcommand{\thefootnote}{\fnsymbol{footnote}}

\newcounter{sectionc}\newcounter{subsectionc}\newcounter{subsubsectionc}
\renewcommand{\section}[1] {\vspace*{0.6cm}\addtocounter{sectionc}{1} 
\setcounter{subsectionc}{0}\setcounter{subsubsectionc}{0}\noindent 
	{\normalsize\bf\thesectionc. #1}\par\vspace*{0.4cm}}
\renewcommand{\subsection}[1] {\vspace*{0.6cm}\addtocounter{subsectionc}{1} 
	\setcounter{subsubsectionc}{0}\noindent 
	{\normalsize\it\thesectionc.\thesubsectionc. #1}\par\vspace*{0.4cm}}
\renewcommand{\subsubsection}[1]
{\vspace*{0.6cm}\addtocounter{subsubsectionc}{1}
	\noindent {\normalsize\rm\thesectionc.\thesubsectionc.\thesubsubsectionc. 
	#1}\par\vspace*{0.4cm}}

\newcounter{appendixc}
\newcounter{subappendixc}[appendixc]
\newcounter{subsubappendixc}[subappendixc]

\renewcommand{\appendix}[1] {\vspace*{0.6cm}
        \refstepcounter{appendixc}
        \setcounter{figure}{0}
        \setcounter{table}{0}
        \setcounter{equation}{0}
        \renewcommand{\thefigure}{\Alph{appendixc}.\arabic{figure}}
        \renewcommand{\thetable}{\Alph{appendixc}.\arabic{table}}
        \renewcommand{\theappendixc}{\Alph{appendixc}}
        \renewcommand{\theequation}{\Alph{appendixc}.\arabic{equation}}
        \noindent{\bf Appendix \theappendixc #1}\par\vspace*{0.4cm}}

\def\abstracts#1{{
	\centering{\begin{minipage}{12.2truecm}\footnotesize\baselineskip=12pt\noindent
	\centerline{\footnotesize ABSTRACT}\vspace*{0.3cm}
	\parindent=0pt #1
	\end{minipage}}\par}} 

\renewenvironment{thebibliography}[1]
	{\begin{list}{\arabic{enumi}.}
	{\usecounter{enumi}\setlength{\parsep}{0pt}
\setlength{\leftmargin 1.25cm}{\rightmargin 0pt}
	 \setlength{\itemsep}{0pt} \settowidth
	{\labelwidth}{#1.}\sloppy}}{\end{list}}

\newcounter{itemlistc}
\newcounter{romanlistc}
\newcounter{alphlistc}
\newcounter{arabiclistc}

\newcommand{\fcaption}[1]{
        \refstepcounter{figure}
        \setbox\@tempboxa = \hbox{\footnotesize Fig.~\thefigure. #1}
        \ifdim \wd\@tempboxa > 6in
           {\begin{center}
        \parbox{6in}{\footnotesize\baselineskip=12pt Fig.~\thefigure. #1}
            \end{center}}
        \else
             {\begin{center}
             {\footnotesize Fig.~\thefigure. #1}
              \end{center}}
        \fi}

\newcommand{\tcaption}[1]{
        \refstepcounter{table}
        \setbox\@tempboxa = \hbox{\footnotesize Table~\thetable. #1}
        \ifdim \wd\@tempboxa > 6in
           {\begin{center}
        \parbox{6in}{\footnotesize\baselineskip=12pt Table~\thetable. #1}
            \end{center}}
        \else
             {\begin{center}
             {\footnotesize Table~\thetable. #1}
              \end{center}}
        \fi}

\def\@citex[#1]#2{\if@filesw\immediate\write\@auxout
	{\string\citation{#2}}\fi
\def\@citea{}\@cite{\@for\@citeb:=#2\do
	{\@citea\def\@citea{,}\@ifundefined
	{b@\@citeb}{{\bf ?}\@warning
	{Citation `\@citeb' on page \thepage \space undefined}}
	{\csname b@\@citeb\endcsname}}}{#1}}

\newif\if@cghi
\def\cite{\@cghitrue\@ifnextchar [{\@tempswatrue
	\@citex}{\@tempswafalse\@citex[]}}
\def\citelow{\@cghifalse\@ifnextchar [{\@tempswatrue
	\@citex}{\@tempswafalse\@citex[]}}
\def\@cite#1#2{{$\null^{#1}$\if@tempswa\typeout
	{IJCGA warning: optional citation argument 
	ignored: `#2'} \fi}}

 1
 1
 1

\font\ninerm=cmr9

\def\alt{\mathrel{\mathop
  {\hbox{\lower0.5ex\hbox{$\sim$}\kern-0.8em\lower-0.7ex\hbox{$<$}}}}}
\def\agt{\mathrel{\mathop
  {\hbox{\lower0.5ex\hbox{$\sim$}\kern-0.8em\lower-0.7ex\hbox{$>$}}}}}

\begin{document}

\vbox{\vskip-48pt\centerline{\fbox{\parbox{\textwidth}
{\noindent To be published in
Proc.\ New Trends in Neutrino Physics,
24--29 May 1998, Ringberg Castle, Tegernsee, Germany,
edited by B.~Kniehl, G.~Raffelt and N.~Schmitz.}}}
\vskip36pt}

\vskip-12pt

\centerline{\normalsize\bf SOME ASTROPHYSICAL IMPLICATIONS OF 
EXPERIMENTALLY}
\baselineskip=16pt
\centerline{\normalsize\bf FAVORED NEUTRINO MASSES AND MIXINGS}

\vspace*{0.6cm}

\centerline{\footnotesize GEORG G.~RAFFELT}
\baselineskip=13pt
\centerline{\footnotesize\it Max-Planck-Institut f\"ur Physik
(Werner-Heisenberg-Institut)}
\baselineskip=12pt
\centerline{\footnotesize\it F\"ohringer Ring 6, 80805 M\"unchen,
Germany}
\centerline{\footnotesize E-mail: raffelt@mppmu.mpg.de}

\vspace*{0.9cm} 

\abstracts{Positive evidence for neutrino oscillations is mounting
  from the solar and atmospheric neutrino anomalies and the LSND
  experiment. Accordingly, the neutrino mass differences appear to lie
  in the sub-eV range and at least some of the mixing angles appear to
  be large. We explore some of the consequences of this picture for
  supernova (SN) physics and cosmology. In particuler, we discuss if
  a conflict between the solar vacuum solution and the SN~1987A signal
  can be avoided, and the role of future cosmic microwave background
  measurements to reveal the presence of a cosmological hot dark
  matter component.}
 
\vspace*{0.6cm}

\normalsize\baselineskip=15pt
\setcounter{footnote}{0}
\renewcommand{\thefootnote}{\alph{footnote}}


\section{Introduction}

The impressive up-down asymmetry observed at
Superkamiokande\cite{SuperKam} leaves little room for doubt that the
atmospheric neutrino anomaly is indeed caused by oscillations of muon
neutrinos into nearly maximally mixed tau or sterile neutrinos.
Likewise, it has become more and more difficult to avoid neutrino
oscillations as an explanation for the solar neutrino
puzzle\cite{Solar}. Only the LSND evidence for
$\bar\nu_e$-$\bar\nu_\mu$-oscillations\cite{LSND} has come under
increasing pressure from the advancing exclusion limit of the KARMEN
experiment\cite{KARMEN}, but thus far remains a viable hypothesis.  

These results (Table~\ref{tab:masses}) suggest three plausible,
qualitatively different scenarios. If LSND is wrong, solar and
atmospheric neutrinos alone indicate very small differences between
the neutrino masses. We may then have a hierarchical mass scheme with
the largest mass eigenvalue below $0.1~\rm eV$.  Or else we may have a
degenerate scheme where all three masses are large relative to their
splittings. Either way, at least some (and perhaps all) of the mixing
angles are large.  If the LSND signal is caused by oscillations after
all, we need a sterile neutrino in addition to the three sequential
ones\cite{fourflavor}. It is then most natural to expect a mass scheme
with two small and two nearly degenerate ``large'' eigenstates,
where the separation between the two pairs is given by LSND. Within
the pairs we can have large mixing.

If we take any of these scenarios seriously, what does this imply for
the possible role of neutrinos in astrophysics and cosmology?  Apart
from the direct experimental observations of solar and atmospheric
neutrinos, the classic astrophysical environments where massive
neutrinos and their oscillations can play a crucial role are
supernovas and the early universe.  Consequently, we shall explore
some of the implications of the experimentally favored neutrino masses
and mixings for SN physics and cosmology.

\begin{table}[t]
\tcaption{Experimentally favored neutrino mass differences and
mixing angles.\label{tab:masses}}
\smallskip
\hbox to\hsize{\hss\vbox{\hbox{\begin{tabular}[4]{llll}
\hline
\noalign{\vskip2pt}
\hline
\noalign{\vskip2pt}
Experiment&Favored Channel&$\Delta m^2$ [$\,\rm eV^2$]&$\sin^22\theta$\\
\noalign{\vskip2pt}
\hline
\noalign{\vskip2pt}
LSND$^a$
&$\bar\nu_\mu\to\bar\nu_e$&0.2--10&(1--$30)\times10^{-3}$\\
Atmospheric&$\nu_\mu\to\nu_\tau$ or $\nu_s$&$10^{-3}$--$10^{-2}$
&0.8--1\\
Solar                             \\
\quad Vacuum&$\nu_e\to{}$anything&$(0.5$--$8)\times10^{-10}$&0.5--1\\   
\quad MSW (small angle)&$\nu_e\to{}$anything&(0.4--1)${}\times10^{-5}$
&$10^{-3}$--$10^{-2}$\\
\quad MSW (large angle)$^b$
&$\nu_e\to\nu_\mu$ or $\nu_\tau$&
(0.3--3)${}\times10^{-4}$&0.6--1\\
\noalign{\vskip2pt}
\hline
\noalign{\vskip2pt}
\hline
\multicolumn{4}{l}{$^a$In near-conflict with 
KARMEN\protect\cite{KARMEN}.}\\
\multicolumn{4}{l}{$^b$Not viable with Superkamiokande spectral
information\protect\cite{Solar}.}\\
\end{tabular}}}\hss}
\end{table}


\section{Cosmic Background Neutrinos}

It is exactly twentyfive years ago that Cowsik and McClelland put
forth the idea that cosmic background neutrinos, if they had a small
mass, could provide the dark matter of the
universe\cite{Cowsik}. However, it was quickly realized that low-mass
collisionless particles do not easily cluster on small scales. A
simple phase-space argument\cite{TremaineGunn}
(``Tremaine-Gunn-limit'') reveals that neutrino masses would have to
exceed about 20--$30~{\rm eV}$ to provide the dark matter of spiral
galaxies, and 100--$200~{\rm eV}$ for dwarf galaxies. On the other
hand, the age of the universe exceeds about $12\times10^9$~years which
sets an upper limit to the cosmic matter density, translating into an
upper limit of about 30--$40~{\rm eV}$ for the sum of all neutrino
masses\cite{KolbTurner}. Evidently, neutrinos cannot be the dark
matter on all scales even though one might imagine that, by a squeeze,
they could still be the dark matter in spiral galaxies. However, if we
take the experimental indications for oscillations seriously, the
neutrino mass differences are small on the scale of 10~eV. Even if we
use the largest LSND-favored $\Delta m^2$ of about $10~{\rm eV}^2$, it
corresponds to $\Delta m\approx 0.5~\rm eV$ for $m\approx 10~\rm
eV$. Therefore, all neutrino masses, including that of a putative
sterile state, are nearly degenerate so that the cosmological limit is
$m\alt10$--15~eV {\em for each flavor separately}, exacerbating the
galactic phase-space problem for neutrino dark matter.

Today it has become difficult to avoid the standard theory of
cosmological structure formation where an essentially flat power
spectrum of primordial density fluctuations is amplified by the
gravitational instability mechanism\cite{KolbTurner}.  Low-mass
neutrinos would stream freely for a long time after their decoupling
in the early universe (``hot dark matter''), erasing small-scale
density perturbations and thus preventing structures to form below the
supercluster scale. Again, low-mass neutrinos do not seem suitable as
dark-matter candidates.

Standard cold dark matter (SCDM) fares much better, but it has the
opposite problem of overproducing small-scale structure. One
intruiging remedy is to invoke a small fraction (say 20\%) of hot dark
matter in a ``hot plus cold dark matter'' (HCDM)
scenario\cite{Primack}.  The small-scale power spectrum of the cosmic
matter-density fluctuations will be measured with unprecedented
precision by the Sloan Digital Sky Survey\cite{sloan}. It was recently
shown that these measurements may well be sensitive down to the lower
end of LSND-inspired neutrino masses\cite{Hu}.  Even presently
available data seem to favor a HCDM cosmology not only over SCDM, but
also over a variety of modified CDM scenarios such as an undercritical
(open) universe (OCDM), a tilted primordial fluctuation spectrum
(TCDM), or CDM with a cosmological constant
($\Lambda$CDM)\cite{Gawiser}. On the other hand, the Hubble diagram of
type Ia supernovae, which was recently extended to high redshifts,
appears to favor a significant $\Lambda$-term and thus a
$\Lambda$CDM~universe\cite{Lambda}.

A cosmological role for neutrinos as a cosmological hot dark matter
component in a HCDM scenario fits in with two of the neutrino mass
schemes discussed in the introduction. We may have three nearly
mass-degenerate neutrino states with a common mass scale of, say,
1--2~eV. This possibility implies that $\nu_e$ has a mass in this
range, making it potentially accessible to searches in neutrinoless
$\beta\beta$ and tritium $\beta$-endpoint experiments.  However, a
truly positive indication for eV-range neutrino masses is provided
only by LSND so that the HCDM hypothesis is closely intertwined with a
four-flavor picture involving sterile neutrinos. They would be
partially excited in the early universe by oscillations from those
sequential neutrinos with which they are mixed so that there would be
a nonstandard radiation component in addition to the hot dark matter
contribution\cite{BBN}. Standard big-bang nucleosynthesis arguments
already constrain an additional radiation content equivalent to more
than one extra neutrino species so that a sterile neutrino which
solves the atmospheric neutrino anomaly by $\nu_\mu$-$\nu_s$
oscillations is already difficult to accomodate.

Besides the impact on the large-scale structure of the universe, a hot
dark matter component would leave its imprint in the temperature
variations of the cosmic microwave background radiation
(CMBR)\cite{Ma}.  The anticipated sky maps of the future MAP and
PLANCK\cite{MAP+PLANCK} satellite missions have already received
advance praise as the ``Cosmic Rosetta Stone''\cite{BTW} because of
the wealth of cosmological precision information they are expected to
reveal\cite{CMBreview}.

\begin{figure}[t]
\epsfxsize=8cm
\hbox to\hsize{\hss\epsfbox{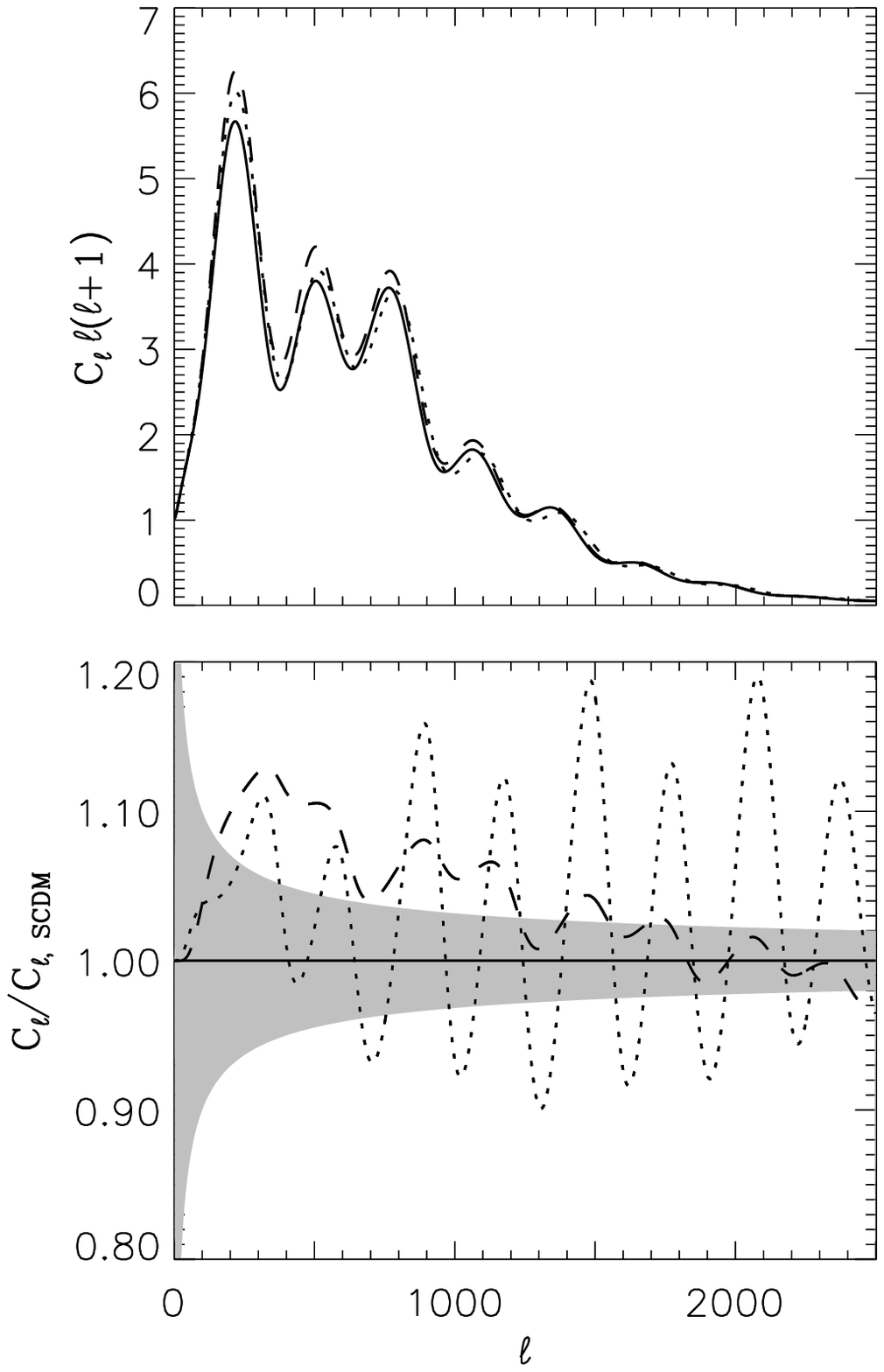}\hss}
\fcaption{{\it Top:} CMBR fluctuation spectrum for SCDM with $h=0.5$,
  $\Omega_M=1$, $\Omega_B=0.05$, and $N_{\rm eff}=3$ (solid line). The
  dotted line is for $N_{\rm eff}=4$, and the dashed line when two of
  these four neutrinos have equal masses corresponding together to
  $\Omega_{\rm HDM}=0.1$ ($\Omega_{\rm CDM}=0.85$).  {\it
  Bottom:}~Relative difference of these nonstandard models to SCDM.
  The shaded band represents the cosmic variance.
  (Figure from Hannestad and Raffelt\cite{Hannestad}.)}
\label{fig:cmb}
\end{figure}

CMBR sky maps are characterized by their fluctuation spectrum
$C_\ell=\langle a^{}_{\ell m} a^*_{\ell m}\rangle$ where $a_{\ell m}$
are the coefficients of a spherical-harmonic
expansion. Figure~\ref{fig:cmb} (solid line) shows $C_\ell$ for
standard cold dark matter (SCDM) with $h=0.5$ for the Hubble constant
in units of $100~{\rm km~s^{-1}~Mpc^{-1}}$, $\Omega_M=1$ and
$\Omega_B=0.05$ for the matter and baryon content, a
Harrison-Zeldovich spectrum of initial density fluctuations, ignoring
reionization, and taking $N_{\rm eff}=3$ for the effective number of
thermal neutrino degrees of freedom.  As a dotted line the power
spectrum is shown for $N_{\rm eff}=3$, and as a dashed line if two of
them carry a small mass such as to contribute 10\% of hot dark
matter\cite{Hannestad}. In the lower panel, the relative difference to
the standard case is shown, together with the ``cosmic variance''
(shaded region) which gives us a rough estimate of the best one can
hope to achieve with the forthcoming experiments. We probably have to
wait for these precision experiments before the presence or absence of
neutrino dark matter, with or without an extra radiation component,
can be finally settled.

In summary, it is conceivable, and even favored by current
cosmological data, that neutrinos with eV-masses play a role as a hot
dark matter component. This hypothesis fits nicely with the LSND
result which is, however, under attack from the KARMEN non-observation
of oscillations. If LSND is eventually proven wrong we are still left
with the possibility of a degenerate three-flavor mass scheme for
which one would not have to worry about extra contributions to the
cosmic radiation density.  It is also possible, however, that
neutrinos do have small masses, but that they obey a hierarchy with
the largest scale below 0.1~eV as suggested by the atmospheric
neutrino anomaly. It would be the ultimate irony if the
Superkamiokande announcement that neutrinos do oscillate, twentyfive
years after Cowsik and McClelland's seminal paper\cite{Cowsik}, would
turn into the beginning of the end of neutrino dark matter!


\section{Supernova Neutrinos}

As the oscillation interpretation of the atmospheric neutrino anomaly
was put onto firmer ground by the Superkamiokande
results\cite{SuperKam}, the perception of the solar neutrino anomaly
has also changed. First, the phenomenon of oscillations has simply
become more real and acceptable. Second, the large mixing angle needed
for atmospheric neutrinos has propelled the solar vacuum solution from
a somewhat obscure also-ran to a bit of a frontrunner.  Variations of
bi-maximal mixing scenarios\cite{bimax} are very much en vogue these
days!

\begin{figure}[b]
\hbox to\hsize{\hss\epsfbox{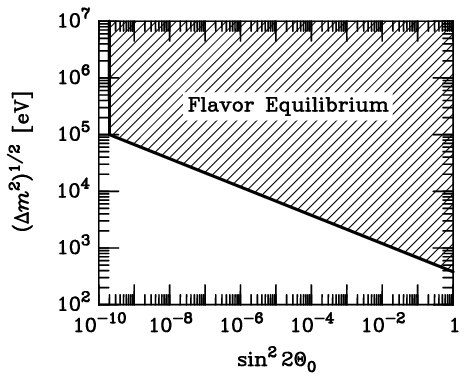}\hss}
\fcaption{In the hatched range of mass differences and
vacuum mixing angles, flavor equilibrium between $\nu_e$ and
$\nu_\mu$ or $\nu_\tau$ would be achieved in a SN core within
$1~{\rm s}$ of collapse\protect\cite{RaffeltBook}.}
\label{fig:conversion}
\end{figure}

The assumption of essentially maximal mixing between $\nu_e$ and the
two other flavors could have important ramifications for supernova
physics. Immediately after the core collapse of a massive star at the
end of its life, neutrinos are trapped, which implies that the
electron-lepton number is almost as large as that of the progenitor's
iron core of about $Y_e=0.35$ electrons per baryon, implying 
highly degenerate Fermi seas of electrons and electron neutrinos. 
The lepton number is thought to be lost on a diffusion timescale
of about $1~{\rm s}$. One might think that the large electron-lepton
number violation implied by maximal $\nu_e$ mixing with $\nu_\mu$ and
$\nu_\tau$ would quickly lead to a redistribution of the 
electron-lepton number among all flavor, creating degenerate seas
of muons and all neutrino flavors. 

However, this is not so because the usual ``quantum-damping'' of the
flavor ``polarization'' by oscillations and collisions is proportional
to $\sin^22\theta$ and the collision rate\cite{Stodolsky}. The large
difference between the refractive index of $\nu_e$ and the other
flavors caused by the medium ``de-mixes'' them enough to slow down the
flavor conversion very much, i.e.\ the in-medium mixing angle becomes
very small\cite{Conversion}.  Before the lepton number is lost by
diffusion within about $1~{\rm s}$, flavor equilibrium is reached only
if the mixing parameters lie in the hatched region of
Fig.~\ref{fig:conversion}. Thus, the matter effect prevents maximal
neutrino mixing from playing havoc with the neutrino flavors in a SN
core as long as the neutrino masses obey the cosmological mass limit.

Another issue arises in the context of the interpretation of the
SN~1987A neutrino signal\cite{Kam,IMB}. It is usually thought that a
SN emits the binding energy of the newly formed compact
star\cite{Janka1},
\begin{equation}
E_{\rm b}= 1.5{-}4.5\times10^{53}\,{\rm erg},
\label{E001x}
\end{equation}
almost entirely in the form of
neutrinos. It is thought that the energy is equally distributed
among the flavors, but with different
spectra which are characterized by the average energies\cite{Janka2}
\begin{equation}
\langle E_{\nu}\rangle=\cases{10{-}12\,{\rm MeV}&for $\nu_e$,\cr
14{-}17\,{\rm MeV}&for $\bar\nu_e$,\cr
24{-}27\,{\rm MeV}&for $\nu_{\mu,\tau}$ and
$\bar\nu_{\mu,\tau}$,} \label{E001}
\end{equation}
i.e.\ $\langle E_{\nu_e}\rangle\approx \frac{2}{3}\langle
E_{\bar\nu_e}\rangle$ and $\langle
E_{\nu}\rangle\approx\frac{5}{3} \langle E_{\bar\nu_e}\rangle$
for the other flavors. A partial conversion between, say, $\bar
\nu_\mu$ and $\bar\nu_e$ due to oscillations would ``stiffen''
the $\bar\nu_e$ spectrum observable at
Earth\cite{Wolfenstein,SSB}. (We will always take
$\bar\nu_e$-$\bar\nu_\mu$-oscillations to represent either
$\bar\nu_e$-$\bar\nu_\mu$ or
$\bar\nu_e$-$\bar\nu_\tau$-oscillations.) Put another way,
some of the SN~1987A $\bar\nu_e$'s observed at the Kamiokande and IMB
detectors would have been oscillated $\bar\nu_\mu$'s which should have
been correspondingly more energetic.

\begin{figure}[ht]
\epsfxsize=8cm \hbox to\hsize{\hss\epsfbox{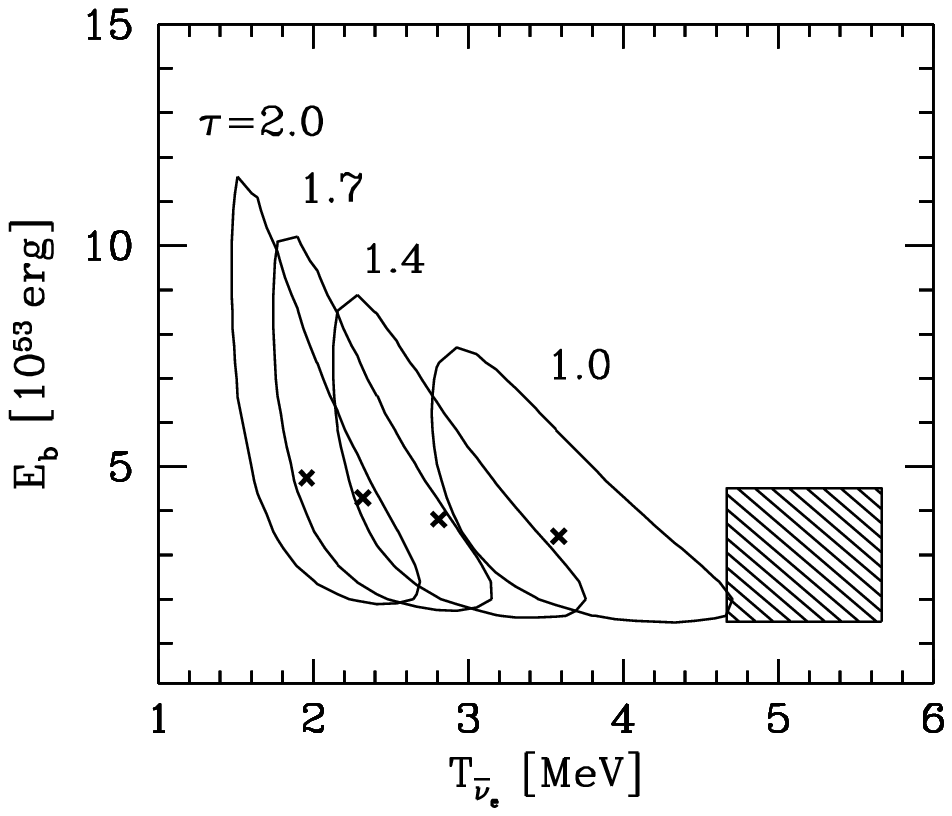}\hss}
\fcaption{Best-fit values for the spectral $\bar\nu_e$ temperature
$T_{\bar\nu_e}$ and the neutron-star binding energy $E_{\rm b}$, as
well as contours of constant likelihood which correspond to 95\%
confidence regions. In each case a joint analysis between the
Kamiokande and IMB detectors was performed with $\sin^22\theta_0=1$
for the vacuum mixing angle and the indicated relative $\bar\nu_\mu$
temperature $\tau$.  The no-oscillation case is equivalent to
$\tau=1$.  The hatched region corresponds to the theoretical
predictions of Eqs.~(\protect\ref{E001x}) and~(\protect\ref{E001}).
[Plot from Jegerlehner, Neubig and Raffelt\protect\cite{Neubig}.]}
\label{fig:sn}
\end{figure}

A maximum-likelihood analysis of the $\bar\nu_e$ spectral temperature
and neutron-star binding energy inferred from the Kamiokande\cite{Kam}
and IMB\cite{IMB} data (Fig.~\ref{fig:sn}) reveals that even in the
no-oscillation case (the $\tau=1$ contour) there is only marginal
overlap with the theoretical expectation of Eqs.~(\protect\ref{E001x})
and (\protect\ref{E001}). Essentially the observed neutrinos were
softer than expected, especially at Kamiokande. Including a partial
swap of the spectra exacerbates this problem in that the expected
energies should have been even higher. In Fig.~\ref{fig:sn} we show
95\% likelihood contours for the inferred $\bar\nu_e$ spectral
temperature $T_{\bar\nu_e}=\langle E_{\bar\nu_e}\rangle/3$ and the
neutron-star binding energy $E_{\rm b}$ for maximum
$\bar\nu_e$-$\bar\nu_\mu$-mixing and for several assumed values
$\tau=T_{\bar\nu_\mu}/T_{\bar\nu_e}$ for the relative spectral
temperature between $\bar\nu_\mu$ and $\bar\nu_e$. Even for moderate
spectral differences a maximum mixing between $\bar\nu_e$ and the
other flavors appears to cause a conflict with the SN~1987A
data\cite{SSB,Neubig}, ostensibly excluding the solar vacuum solution.

However, the SN~1987A data and the solar vacuum solution could well be
compatible if the spectral differences between the different flavors
had been theoretically overestimated in the past. These differences
arise because the opacities are different for the different
flavors---the electron-flavored neutrinos are primarily trapped by
charged-current processes of the type $\nu_e n\leftrightarrow p e^-$
or $\bar\nu_e p\leftrightarrow n e^+$ while the other flavors scatter
by neutral currents mostly on nucleons and thus decouple in deeper
layers at higher temperatures.  The non-electron flavored neutrinos
escape from their ``transport sphere'' where scattering processes are
no longer effective. However, most critical for their spectrum is the
deeper-lying ``energy sphere'' where they last exchange energy with
the medium; the scattering on nucleons was usually treated as being
ineffective at energy transfer because recoils are suppressed by the
large nucleon mass. Therefore, the energy sphere was defined by
electron scattering $\nu e^-\to e^-\nu$ while $e^+e^-\to \nu\bar\nu$
was taken as the dominant production process. 

However, this approximation is strictly incorrect. It was recently
shown\cite{Suzuki,Keil,HR} that the dominant pair-process is nucleonic
bremsstrahlung $NN\to NN\nu\bar\nu$ while the dominant
energy-exchange process is inelastic neutrino-nucleon scattering
$\nu NN\to NN\nu$ and recoils in $\nu N\to N\nu$.  Including these
dominant energy-exchange processes clearly has the effect of making
the $\bar\nu_\mu$ spectrum more similar to the $\bar\nu_e$ one. A
quantitative estimate\cite{HR} suggests that the remaining spectral
differences may be small enough to avoid any conflict between the
SN~1987A data and the solar vacuum solution. Of course, a
self-consistent calculation of the neutrino fluxes and spectra with
the correct microphysics is needed to substantiate this conclusion.

The main point that can be raised at the present time is that the
canonical large spectral differences between the different neutrino
flavors emitted from a SN core are rather questionable and thus cannot
be used for far-reaching conclusions regarding the SN~1987A signal
interpretation.  Likewise, a variety of neutrino-oscillation issues in
SN physics, from r-process nucleosynthesis to pulsar kicks, would be
affected by a significant revision of the flavor-dependence of SN
neutrino spectra.


\end{document}